\RequirePackage[T1]{fontenc}
\documentclass[conference,10pt]{IEEEtran}

\usepackage{amsmath}
\usepackage{algorithm}
\usepackage{algpseudocode}
\usepackage{newtxtext,newtxmath}
\usepackage{booktabs}
\usepackage{graphicx}
\usepackage{microtype}
\usepackage{multirow}
\usepackage{tabularx}
\usepackage{url}

\newcommand{\system}{QEffect}
\newcommand{\tkey}{\ensuremath{\mathsf{TapeKey}}}
\newcommand{\tref}{\ensuremath{\mathsf{TapeRef}}}

\title{Explicit State and Resource Contracts for Low-Precision Pipeline Parallel Training under Captured Graphs}

\IEEEoverridecommandlockouts
\author{\IEEEauthorblockN{Genlang Chen\textsuperscript{*}\thanks{\textsuperscript{*}Corresponding author: cgl@zju.edu.cn}}
\IEEEauthorblockA{NingboTech University\\
cgl@zju.edu.cn}
\and
\IEEEauthorblockN{Junyi Zhu}
\IEEEauthorblockA{Dalian Ocean University\\
junyizhu.cs@gmail.com}}

\begin{document}
\maketitle

\begin{abstract}
Graph capture mechanisms such as CUDA Graphs amortize host launch latency and framework overhead by replaying predetermined execution plans over fixed device virtual addresses. In low-precision floating-point (FP8) pipeline-parallel training, however, these invariant memory buffers correspond to continuously shifting runtime entities: evolving delayed-scaling factors, distinct microbatch activations, and asynchronously deferred backward tasks. Advanced split-backward schedules (e.g., 1F1B and Zero-Bubble) decouple input-gradient ($dI$) from weight-gradient ($dW$) passes to diminish pipeline bubbles, but this separation disrupts traditional stack-disciplined resource lifecycles. Standard computational dataflow graphs encode tensor dependencies while remaining oblivious to internal numerical state transitions, non-LIFO work ownership, and weight-cache lifetimes---frequently inducing latent race conditions and silent numerical corruption during concurrent graph replay. In this paper, we establish an explicit state and resource contract system (\system) that bridges the semantic gap between graph capture and stateful, low-precision pipeline training. \system{} formalizes four foundational runtime invariants: (1)~temporal serialization of hidden quantization state, (2)~generational ownership of retained backward resources, (3)~validity versioning for cached weights across optimizer boundaries, and (4)~bidirectional completion synchronization between caller streams and graph executors. These invariants uniformly govern both eager execution and captured graph replay, while allowing volatile graph-scoped resources to be cleanly recreated upon checkpoint restoration. Leveraging rigorous work-ownership contracts, we further design an affine direct-gradient placement mechanism that completely bypasses intermediate gradient staging. Implemented atop TorchTitan and NVIDIA Transformer Engine, our runtime maintains strict bitwise parity with native full-backward baselines across delayed-scaling rollover events, deterministically intercepts cross-stream ordering anomalies, and guarantees error-free cold restart. Microbenchmarks on NVIDIA H800 GPUs demonstrate that graph capture accelerates transformer layers by 1.82--2.79$\times$ relative to eager execution, while direct gradient placement secures an additional 1.132$\times$ throughput improvement by pruning 96 redundant tensor transfers per rank per step.
\end{abstract}

\section{Introduction}

While computational graphs effectively model tensor dataflow among operators,
modern training runtimes increasingly rely on hidden numerical state transitions
and out-of-order deferred tasks.  This semantic gap becomes acute when
low-precision arithmetic, pipeline scheduling, and graph replay intersect.
Delayed scaling dynamically adjusts scaling factors to keep FP8 arithmetic
numerically stable~\cite{micikevicius2022fp8,peng2023fp8lm}.  Pipeline runtimes
split backward passes into separate input-gradient ($dI$) and weight-gradient
($dW$) stages to minimize pipeline bubbles~\cite{narayanan2021megatron,qi2024zerobubble}.
CUDA Graphs eliminate Python interpreter and kernel-launch overheads by
replaying fixed execution plans over static addresses~\cite{ansel2024pytorch2,nvidia2026cudagraphs,ghosh2026grace}.
While each technique is sound within its isolated domain, their runtime interfaces
fail to jointly specify hidden numerical state, asynchronous lifetimes, and retained work.

NVIDIA Transformer Engine (TE) exemplifies this interface mismatch.  An
individual layer encapsulates absolute-maximum (amax) histories, dynamic scaling
factors, and quantized-weight caches behind an opaque tensor interface.  Its
backward pass also retains intermediate activations intended for subsequent
weight-gradient evaluation.  Conventional autograd engines rely on strict
last-in, first-out (LIFO) stack discipline to associate retained work with the
corresponding forward execution.  Zero-Bubble pipelining breaks this assumption:
multiple microbatches remain concurrently active, $dI$ updates shared scaling
state, and the matching $dW$ may execute out of order after an arbitrary
schedule delay.  CUDA Graph replay introduces further complications: static
memory addresses are reused while their underlying logical microbatches and
weight versions continuously evolve.

Consequently, an execution graph can appear structurally complete under pure
tensor dataflow while omitting four indispensable relations for correct runtime
execution.  Specifically, the runtime must capture the topological sequence of
hidden numerical updates, the explicit ownership and lifetime of retained work,
the freshness of cached weights, and the completion of asynchronous device
streams.  These orthogonal properties cannot be collapsed into a monolithic
dependency token.  A scalar token enforces sequential execution order but
cannot identify a deferred resource; an identifier names a resource but cannot
order updates to shared scaling state.

Compiler intermediate representations (IRs) traditionally employ tokens and
abstract resource types to sequence side effects
~\cite{openxla2026stablehlo,jax2024effects,mlir2026sideeffects}.  NVIDIA enables
persistent FP8 buffers and weight caching for CUDA Graphs under restricted
Megatron schedules~\cite{nvidia2026transformerenginegraphs}.  GraCE broadens
graph coverage via parameter indirection and selective capture
~\cite{ghosh2026grace}, while Zero Bubble defines split-backward pipelining
~\cite{qi2024zerobubble}.  While foundational, none of these systems reconciles
hidden scaling state, non-LIFO retained work, cache versioning, stream
synchronization, and failure recovery within a unified stage interface.

We designate our approach \system{} because it elevates quantization-related side
effects---including FP8 delayed-scaling state updates, quantized-weight cache
invalidations, and deferred low-precision gradient tasks---into first-class,
verifiable runtime contracts.  \system{} enforces strict ordering on hidden
numerical updates and binds each retained backward resource to a unique logical
action and physical generation.  It tracks cache validity across optimizer
epochs and establishes bidirectional synchronization between caller and graph
streams upon every replay.  Quiescent checkpointing serializes logical and
numerical state only when all in-flight work has settled, enabling ephemeral
graph, stream, and tape objects to be rebuilt cleanly upon restart.  For TE
TransformerLayers, an affine direct-placement optimization exploits
retained-work ownership to write delayed matrix gradients directly into stable
per-microbatch memory arenas, eliminating high-volume memory copies without
altering numerical results.

\begin{figure*}[t]
  \centering
  \includegraphics[width=\textwidth]{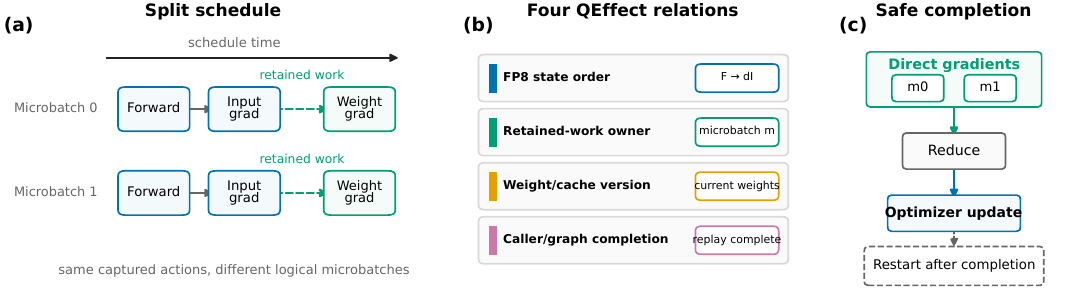}
  \caption{How \system{} preserves a split pipeline action across schedule
  delay and graph replay.  State order, retained-work ownership, weight version,
  and device completion connect each logical action to the correct state and
  destination.  The detailed mechanisms are a device token, \tkey{}/\tref{},
  weight/cache versions, and caller--graph events.}
  \label{fig:model}
\end{figure*}

Figure~\ref{fig:model} illustrates how these contracts compose across stages.

The work makes three principal contributions:

\begin{itemize}
  \item We identify and formalize four relations that tensor dataflow graphs
  omit in captured, split-backward pipelines: state order, retained-work
  ownership, weight version, and device completion.  In an unconstrained
  baseline path, 200 producer--consumer checks across three independent
  processes consistently reveal stale reads in all 600 trials.
  \item We design and implement the \system{} contract in TorchTitan GraphPP.
  The runtime supports multiple outstanding microbatches, non-LIFO $dW$
  scheduling, static-address graph replay, and robust checkpoint resumption from
  quiescent optimizer boundaries.
  \item We comprehensively validate \system{} against native-TE full-backward
  references through fault injection, checkpoint/resume verification, causal
  Nsight profiling, resource scaling, and pipeline mapping studies.  Captured TE
  paths are 1.82--2.79$\times$ faster than native eager execution, while direct
  gradient placement delivers an additional 1.138$\times$ speedup on Ada and
  1.132$\times$ on Hopper.
\end{itemize}

\section{Background and Motivation}
\label{sec:motivation}

\subsection{Temporal Latency and Hidden State in Delayed Scaling}

FP8 formats trade exponent and mantissa range for computational throughput and
memory bandwidth~\cite{micikevicius2022fp8}.  To avoid underflow and saturation,
delayed scaling computes scaling factors from moving absolute-maximum (amax)
histories rather than synchronously inspecting every operand tensor.  A module
invocation therefore reads persistent scaling state, executes quantized matrix
multiplication, and updates internal statistics for subsequent microbatches.
Quantized weights may additionally be cached across microbatch boundaries and
refreshed only during the initial forward pass following an optimizer step.
Consequently, the conventional tensor operator signature $Y=f(X,W)$ completely
omits both the internal numerical state transition and the validity invariants of
cached weight representations.

NVIDIA's CUDA Graph integration makes these internal buffers persistent and
serializes deferred updates for supported Megatron-LM schedules
~\cite{nvidia2026transformerenginegraphs}.  However, split-backward pipelining
demands a strictly stronger interface because its numerical state transitions
and retained gradient computations are scheduled asynchronously.  While static
scaling alternatives circumvent dynamic state updates~\cite{narayan2025unit},
\system{} specifically targets dynamic and delayed scaling.

\subsection{Non-LIFO Lifetimes in Split-Backward Pipelining}

Conventional pipeline schedules treat backward propagation as an atomic stage
action.  Zero-Bubble pipelining observes that input gradients ($dI$) lie on the
critical latency path of the pipeline bubble, whereas weight gradients ($dW$)
can be safely deferred to bubble regions~\cite{qi2024zerobubble}.  At $dI(k)$,
the runtime reads saved activations from forward microbatch $k$ and queues
intermediate tasks that a subsequent $dW(k)$ must evaluate.  Because arbitrary
microbatches may interleave between $dI(k)$ and $dW(k)$, weight-gradient
evaluations violate standard LIFO deallocation order.  Retained backward state
must therefore carry an explicit, collision-free identity that survives across
graph captures and replayed invocations.

\subsection{Semantic Invalidity Under Asynchronous Stream Execution}

CUDA Graph capture records device kernels on a dedicated stream and replays them
against fixed virtual memory addresses.  In GraphPP, a pipeline stage captured on
an internal graph stream was replayed from caller streams without re-establishing
device-level happens-before edges.  While replay returned stable output tensor
handles to the host runtime, subsequent graphs or kernels could access shared
state buffers and weight caches before prior device replays finalized.
Parameters and delayed-scaling states silently diverged across microbatches even
while loss curves remained finite.

\begin{figure*}[t]
  \centering
  \includegraphics[width=\textwidth]{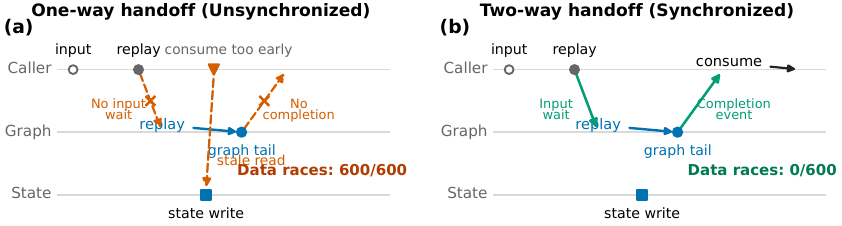}
  \caption{Caller-stream safety requires bidirectional synchronization.  The
  unsynchronized one-way path returns control while a downstream consumer reads
  state before replay completes (producing data races in all 600 trials).  In the
  synchronized two-way path, the graph stream waits for caller input and the
  caller waits for graph completion (eliminating all 600 data races).}
  \label{fig:stream}
\end{figure*}

Figure~\ref{fig:stream} contrasts this defective one-way handoff against
bidirectional stream synchronization.  We validate both protocols across three
independent fresh processes executing 200 producer--consumer checks each.  The
unconstrained one-way protocol triggers data-race hazards and stale reads in all
600 checks because the host proceeds before device replay completes.  In
contrast, the two-way protocol enforces a dual handshake: the graph stream waits
on the incoming caller stream before replay, and the caller stream synchronizes
with a recorded graph-tail event prior to consumption.  This eliminates all
600 data races.

Correct graph replay must therefore preserve runtime semantic contracts and
device completion, not merely tensor shapes and memory pointers.

\subsection{Insufficiency of Monolithic Dependency Tokens}

While an opaque scalar token can serialize execution order, temporal sequencing
addresses only a fraction of the problem.  A deferred pipeline action must also
bind to the exact retained work it consumes, verify that cached parameters
reflect the latest optimizer commit, and synchronize asynchronous device streams.
Collapsing all stage dependencies into a single monolithic token introduces false
serialization dependencies and cripples pipeline concurrency while leaving
resource ownership implicit.

Table~\ref{tab:gap} maps each hidden runtime object to its required contractual
relation.  \system{} unifies state serialization, retained-work ownership,
cache versioning, and stream completion at the stage boundary.  Checkpointing
applies these same relations to cleanly decouple persistent model states from
ephemeral, process-local runtime structures.

\begin{table*}[t]
\centering
\caption{Four relations omitted by ordinary tensor dataflow and the mechanism
that represents each relation in \system{}.}
\label{tab:gap}
\footnotesize
\setlength{\tabcolsep}{4.5pt}
\begin{tabularx}{\textwidth}{@{}p{0.12\textwidth}p{0.18\textwidth}p{0.30\textwidth}>{\raggedright\arraybackslash}X@{}}
\toprule
Relation & Hidden object & Failure without the relation & \system{} mechanism \\
\midrule
Order & FP8 scaling state & duplicated, skipped, or reordered scale update & device token and persistent FP8 state \\
Ownership & retained $dW$ work & wrong microbatch or a reference to a reused slot & logical action identity and generation-checked reference \\
Version & quantized-weight cache & a fixed address represents stale weights & separate weight and cache versions; first-forward refresh \\
Completion & caller and graph streams & a consumer observes a partially updated state & caller-to-graph wait and graph-to-caller completion event \\
\bottomrule
\end{tabularx}
\end{table*}

Checkpointing composes the same relations.  It waits for retained work and
device operations to complete, persists FP8 state and weight versions, and
assigns new process-local identities to rebuilt tapes, events, and caches.

\section{QEffect Contract}
\label{sec:model}

\system{} formalizes a pipeline stage as three disjoint operator actions:
forward ($F$), input-gradient computation ($dI$), and weight-gradient computation
($dW$).  Prior to executing an action, the contract determines which hidden state
transition occurs, which retained backward work belongs to the action, which
parameter version its cache represents, and whether preceding asynchronous device
operations have finalized.  Identical tensor operator signatures and validation
rules govern both eager execution and prebuilt CUDA Graphs.

\subsection{Naming retained work}

Static virtual memory addresses in captured graphs cannot identify dynamic
training iterations because identical physical slots are recycled across
microbatches and optimizer epochs.  \system{} therefore identifies each stage
execution via an affine invocation key:
\begin{equation}
  K = \tkey(e,s,m,u,i),
\end{equation}
where $e$ denotes the optimizer epoch, $s$ the pipeline stage, $m$ the
microbatch index, $u$ the sub-layer block (e.g., attention or MLP), and $i$ an
intra-module invocation counter.  This logical identifier is strictly decoupled
from physical device buffer locations.

Because CUDA Graph execution precludes dynamic tensor allocations during replay,
physical storage is provisioned via a bounded slot pool.  Allocating a tape in
slot $p$ increments a monotonically increasing generation counter $g_p$ and
yields an affine reference:
\begin{equation}
  R = \tref(p,g_p).
\end{equation}
The active mapping is strictly injective: each logical key maps to at most one
live physical tape, and an occupied slot cannot be preempted.  A consuming
operator must present both the logical key and the exact expected generation.
Releasing a slot evicts its tensor payload while preserving the monotonic
progression of $g_p$, provably eliminating use-after-free or ABA-style stale
accesses upon address reuse.

Each tape tracks consumer completion via an internal bitmask.  The pool traps
duplicate consumer invocations and releases physical storage only after both
$dI$ and $dW$ have executed successfully.  Runtime exceptions abort the affected
reference, and an aborted step purges outstanding tapes associated with epoch $e$.

\subsection{Ordering hidden state}

Captured operators sequence hidden numerical effects by threading an explicit
scalar device token.  This token carries no numerical tensors; its sole purpose
is to enforce device-level topological ordering among stateful kernels.  For
microbatch $m$, the state transition is defined by:
\begin{align}
  (Y_m,S_1,T_m) &= F(X_m,W,S_0), \\
  (dX_m,B_m,S_2,Q_m) &= dI(dY_m,T_m,S_1), \\
  dW_m &= dW(Q_m),
\end{align}
where $S$ represents delayed-scaling internal state, $T_m$ is the forward
activation tape, $B_m$ contains non-matrix parameter gradients, and $Q_m$
denotes retained matrix-gradient tasks.  TE quantizes incoming output gradients
and updates the backward amax history exactly once during $dI$.  The subsequent
$dW$ consumes $Q_m$ without re-executing this update.  The effect token therefore
enforces *when* a state transition occurs, while the tape reference determines
*which* microbatch owns the deferred computation.

\subsection{Tracking weight versions}

\system{} maintains a logical optimizer epoch $e_W$ and a quantized-weight cache
epoch $e_C$.  An optimizer commit advances $e_W$ only after all backward tapes,
arena writes, and distributed reductions have finalized.  Every forward action
requires parameter epoch consistency ($e_W$).  The initial microbatch of a step
refreshes the quantized TE cache and synchronizes $e_C \leftarrow e_W$;
subsequent microbatches verify cache freshness against this version.  Following an
optimizer step or process restart, stale caches are automatically refreshed
during the first forward pass.

The captured graph maintains fixed process-local pointers while GraphPP advances
the logical epoch.  Fixed addresses can thus represent successive training
iterations without permitting stale parameter reuse.

\subsection{Optional direct gradient placement}

Surfacing matrix gradients across custom CUDA Graph operator boundaries introduces
an aliasing challenge: TE frequently exposes views into internal accumulation
buffers, whereas custom-operator graph interfaces require independent output
tensors.  A conservative reference implementation clones these gradients, but at
hidden size 4,096, this extraneous memory copy can nullify the latency benefits
of graph capture.

To circumvent this penalty, \system{} allocates a dedicated matrix-gradient arena
providing stable preallocated views for each matrix parameter and configured
microbatch.  Prior to invoking $dW(m)$, the runtime binds these views directly as
the TE \texttt{main\_grad} accumulation targets, enabling fused in-kernel gradient
accumulation.  The custom CUDA Graph operator mutates the prebound arena in place
and returns solely its effect token.  Runtime assertions strictly validate tensor
shape, data type, device ID, and \texttt{data\_ptr} for every bound destination.
Microbatch 0 initializes the accumulation buffer; subsequent microbatches write
according to the pipeline's arbitrary non-LIFO schedule.  Duplicate, missing, or
aliased writes trigger immediate failures.  Finalization mandates that every
configured microbatch writes exactly once prior to optimizer commit.

Direct placement alters only the physical destination of $dW$, leaving the
underlying numerical state machine intact.  The unoptimized copied-gradient path
shares the identical four correctness contracts without utilizing the arena.

\subsection{Restarting after pending work completes}

Checkpoints are serialized exclusively when no active tapes, arena writes, or
distributed reductions remain in flight.  \system{} serializes TE numerical state,
logical optimizer epochs, cache semantics, precision recipes, and static
configurations.  Graph objects, device streams, CUDA events, tape generations,
arena addresses, effect tokens, and cache contents are treated as ephemeral
process-local state and are reconstructed upon restart.  Resumption restores
numerical state prior to graph instantiation and invalidates weight caches to
guarantee a clean refresh on the initial forward pass.

\subsection{Contract invariants}

\system{} enforces four foundational invariants across stage boundaries:

\noindent\textbf{Invariant 1 (Temporal Serialization).} State-mutating forward and input-gradient passes advance an explicit scalar effect token.  Numerical scaling updates are performed strictly once per forward-backward cycle; delayed $dW$ execution never re-executes or perturbs scaling state.

\noindent\textbf{Invariant 2 (Generational Ownership).} Each active $\tkey$ binds to exactly one physical tape slot, guarded by a monotonically incrementing generation counter $g_p$.  Dereferencing verifies generation equality, preventing stale reads and ABA hazards across slot reuse.

\noindent\textbf{Invariant 3 (Cache Freshness).} Forward invocations validate parameter epoch consistency ($e_C = e_W$).  The initial microbatch following an optimizer step forces an in-place weight cache refresh, after which subsequent microbatches verify cache currency.

\noindent\textbf{Invariant 4 (Quiescent Persistence).} Checkpointing occurs exclusively when all active tapes, arena writes, and gradient reductions have resolved.  Persistent numerical state is serialized alongside logical epochs, while all ephemeral device handles are cleanly reconstructed upon restart.

\section{Runtime Integration}
\label{sec:implementation}

We implement \system{} in TorchTitan GraphTrainer and GraphPP
~\cite{liang2025torchtitan}.  The runtime wraps a TE
\texttt{TransformerLayer} with delayed scaling and connects it to the language
model, optimizer, pipeline communication, optional data-parallel reduction,
and PyTorch Distributed Checkpoint (DCP)~\cite{pytorch2026dcp}.

\subsection{Extracting F, $dI$, and $dW$}

F runs the TE layer under FP8 autocast and creates a retained
\texttt{WeightGradStore} for each matrix-operation domain.  It copies dynamic
inputs into fixed graph buffers, gives autograd a distinct retained input, and
binds the input, output, and stores to the slot named by $K$.

$dI$ resolves the corresponding \tref{}, checks its key and epoch, and uses
autograd to compute the input gradient and ten non-matrix parameter gradients.
TE leaves the six matrix gradients as deferred work in four stores.  After all
four stores are ready, $dI$ consumes its part of the tape.  Later, $dW$ resolves
the same generation, drains the stores, and checks that each matrix parameter
is covered exactly once.  The copied-gradient path returns independent gradients;
the direct-placement path checks the bound buffers and returns only the
ordering token.  Thus
all 16 stage parameters are accounted for before the optimizer step.

The three actions are CUDA custom operators with explicit tensor signatures.
One F/$dI$/$dW$ triple is bound to each fixed tape slot and is shared by eager
and CUDA Graph execution.  Algorithm~\ref{alg:protocol} gives the optimizer-step
protocol in schedule order.  A check precedes its physical action, and the
corresponding transition is committed only after that action succeeds.

\begin{algorithm}[t]
\caption{\system{} protocol for one optimizer step.}
\label{alg:protocol}
\footnotesize
\begin{algorithmic}[1]
\Require Pipeline-scheduled actions for the current weight version
\For{each action $a$ in schedule order}
  \If{$a$ is Forward}
    \State Check the weight/cache version and run F
    \State Bind retained work to the logical action; advance FP8-state order
  \ElsIf{$a$ is Input grad}
    \State Check action identity, generation, and weight version
    \State Run $dI$; require delayed $dW$ work; complete the $dI$ ownership
  \ElsIf{$a$ is Weight grad}
    \State Resolve the same generation and run $dW$
    \State If direct placement is enabled, check each bound destination
    \State Complete the $dW$ ownership
  \EndIf
\EndFor
\State Finish reductions and require no pending retained work
\State Require complete direct destinations when enabled; update parameters and version
\Statex \textit{Restart rule:} load FP8 state and versions, rebuild process-local
resources, and refresh the weight cache on the first forward.
\end{algorithmic}
\end{algorithm}

\subsection{GraphPP schedules}

GraphPP already expresses stage forward and split backward.  \system{} replaces
the stage body with the checked actions and carries $(K,R)$ in its saved-value
edge.  Interleaved 1F1B and ZB-V use the same implementation; only the event
order changes.  The logical pool can hold several stage-boundary tapes, and a
$dW$ verifies that both its explicit microbatch index and its reference name
the same slot.  This supports multiple outstanding microbatches and delayed,
non-LIFO $dW$ without relying on a module-global backward stack.

The last stage may use either a stateful TE FP8 output head or a dynamic
TorchAO Float8Linear boundary.  TorchAO exercises the same stage interface,
while the retained delayed-scaling path is implemented and evaluated with TE.

\subsection{Static binding and resource cost}

Prebuilding actions turns dynamic schedule identities into a bounded mapping.
For $M$ microbatches and $L_r$ local stage chunks on rank $r$, the core creates
one graph for each F/$dI$/$dW$ action,
\begin{equation}
  G_r = 3M L_r.
\end{equation}
For $M=2,4,8,16$ and two local chunks, this gives the measured 12, 24, 48,
and 96 graphs.  These counts describe provisioned capacity.  A schedule may
keep far fewer tapes live at once.

The matrix arena has the corresponding capacity
\begin{equation}
  A_r = M \sum_{j \in \mathcal{W}_r} \mathrm{sizeof}(W_j),
\end{equation}
where $\mathcal{W}_r$ contains the local matrix parameters.  The arena is both
reserved storage and the optimizer-visible gradient destination.  It uses
predictable memory to remove repeated full-gradient copies while keeping every
address fixed for capture.

This static binding does not require the schedule to execute all $dW$ actions
in slot order.  The stage initializes accumulation with microbatch 0, as
required by TE's fused accumulation path, then accepts the remaining
microbatches in the schedule order.  Logical identity and generation checks
remain active even though the selected callable is indexed by a static slot.

\subsection{Capture and replay}

Warmup, dynamic-input copies, capture, and replay all run on one manager-owned
graph stream.  At every invocation the graph stream waits on the current caller
stream.  An external CUDA event is recorded at the graph tail during capture,
and the caller waits on it after every replay.  Static parameters and tape-slot
buffers retain their addresses; dynamic input tensors are copied only after the
caller-to-graph edge.  The implementation records capture host time, graph
count, replay count, caller/graph handoffs, arena capacity, and persistent
memory growth so setup and steady-state costs remain distinct.

\subsection{Optimizer and failure boundaries}

The optimizer commit establishes a cluster-wide consistency barrier.  GraphPP
verifies that all backward tapes, arena destinations, and asynchronous data-parallel
reductions are completely resolved before parameter updates occur.  The split
module, output projection head, gradient arena, and gradient reducer then advance
synchronously to the subsequent logical epoch.  If any action encounters an
exception or validation failure, outstanding tapes are immediately aborted and
partially updated arenas are flagged unusable for that epoch.

\section{Evaluation}
\label{sec:evaluation}

We first check whether \system{} preserves numerical and runtime state across
split backward, graph replay, and fresh-process restart, and compare its
semantics with the official TE and Megatron paths.  We then measure complete
optimizer steps, isolate the effect of direct gradient placement, and study
setup, memory, resource capacity, and pipeline mapping.

\subsection{Methodology}

\textbf{Hardware and software.}
Performance runs use two or four NVIDIA H800 PCIe GPUs and either two or four
RTX 5880 Ada GPUs.  All directly compared ranks have peer-to-peer
communication.  The common stack is a source PyTorch build
(\texttt{2.15.0a0+git411c2b5}), CUDA 12.8, TE 2.18, and the NVIDIA Collective
Communications Library (NCCL).  The dynamic
output-head experiment uses TorchAO 0.18 development code.  Every directly
compared configuration loads the same serialized parameter state and data
sequence.

\textbf{Workloads.}
The evaluation examines configurations with hidden, feed-forward, and sequence
dimensions of 2,048 / 8,192 / 256 across four microbatches, and 4,096 / 11,008 / 256
across eight microbatches.  We denote pipeline, data, and virtual pipeline-parallel
degrees by PP, DP, and VPP, respectively.  The primary performance cells deploy
$\mathrm{PP}=2$ with four virtual stage chunks.  Both configurations contain four
\texttt{TransformerLayers} (one per virtual stage), 16/32 attention heads, and a
32,768-token vocabulary.  The scaling benchmark holds the four-stage hidden-4,096
model and 2,048 global tokens constant while transitioning from two local chunks
per rank at $\mathrm{PP}=2$ to one chunk at $\mathrm{PP}=4$.  Unless noted otherwise,
microbatch size is one with sequence length 256.  We intentionally evaluate this
fine-grained, overhead-dominated regime as an adversarial worst-case stress test:
in compute-bound settings with long sequences, runtime and launch overheads are
masked by Tensor Cores; conversely, fine-grained pipeline chunks accentuate graph
launch latencies, stream handoff stalls, and memory aliasing bottlenecks.  Training
uses AdamW on a pinned 2,000-record C4 asset~\cite{raffel2020t5} (1,600 training and
400 held-out records); synthetic inputs are used for bit-exact state verification.
The resource analysis varies microbatches across $\{2, 4, 8, 16\}$ under a fixed
model shape.

\textbf{Timing and statistics.}
The primary metric is the maximum optimizer-step wall time across ranks after
lazy initialization, warmup, compilation, capture, and optimizer-state
materialization.  A fresh process is one trial.  We report the initial trial
set ($n=3$) and an independent replication set ($n=2$) separately using
medians.
Setup-to-break-even uses paired initialization-plus-first-step and steady-state
measurements; instrumented profiler runs are excluded from timing.

\textbf{Baselines.}
Native TE eager uses Interleaved 1F1B with full autograd backward.  The
\system{} eager path and the captured Interleaved paths with copied gradients
or direct placement use the same split F/$dI$/$dW$ schedule.  Captured ZB-V
uses direct placement and changes only the schedule order.  All absolute-time
comparisons between \system{} and native TE
share model state, data, optimizer, precision recipe, and complete-step
boundary.  We also audit TE
\texttt{make\_graphed\_callables} and a pinned native Megatron-LM revision.
Megatron performance is reported as a within-framework graph/eager ratio.

\textbf{State comparison.}
Each shape/backend pair uses a stage-indexed parameter snapshot that loads
identical tensors independently of the rank-to-stage mapping.  Data, FP8
recipe, optimizer, learning rate, input placement, warmup, and stopping rule
are fixed within each comparison.  We compare loss, parameters, gradients,
optimizer state, FP8 scaling state, cache versions, resource counts, and persistent memory
growth.

\textbf{Reproducibility records.}
Schema-checked JSON reports retain the configuration, raw measurements, and
resource counts used by every figure and table.  The artifact includes these
reports, validation scripts, and rendering inputs.

\subsection{Semantic capability and correctness}

Table~\ref{tab:capability} reports only behavior exercised by the audits and
experiments.  Native TE eager is the matched numerical and complete-step
reference.  TE's standalone graphed full backward passes at module scope, but
its integrated PP path produces nonfinite first-step gradients and is excluded
from timing.  Native Megatron's TE graph path runs under PP=2/VPP=2 delayed
scaling.  Its delayed weight-gradient option requires mixture-of-experts (MoE)
overlap, so the dense PP/VPP request stops before training.

\begin{table*}[t]
\centering
\caption{Observed scope and role of the baseline and \system{} paths.  The
Megatron delayed-$dW$ row records the dense PP/VPP request that stops before
training.}
\label{tab:capability}
\footnotesize
\setlength{\tabcolsep}{3.4pt}
\begin{tabularx}{\textwidth}{@{}>{\raggedright\arraybackslash}p{0.14\textwidth}>{\raggedright\arraybackslash}p{0.13\textwidth}>{\raggedright\arraybackslash}p{0.19\textwidth}>{\raggedright\arraybackslash}p{0.28\textwidth}>{\raggedright\arraybackslash}X@{}}
\toprule
Path & Graph scope & Backward schedule & Observed result & Role in evaluation \\
\midrule
Native TE eager & Eager; no graph & Full autograd backward & Completes the matched correctness and timing runs & Numerical and complete-step reference \\
Standalone TE module graph & Standalone module; PP integration probed & Full backward in the standalone probe & Standalone probe passes; PP integration has nonfinite first-step gradients & Compatibility control \\
Megatron TE graph & PP=2/VPP=2 & Full backward & Loss matches eager; 1.751--2.191$\times$ within Megatron & Official graph control \\
Megatron delayed $dW$ & Dense PP/VPP request & Delayed $dW$ requested & Rejected before training; requires MoE overlap & Scope control \\
\system{} Interleaved & PP=2/4 stage graphs & Split F/$dI$/$dW$; schedule-defined $dW$ & Zero state mismatches; exact restart & Evaluated system \\
\system{} ZB-V & PP=2 stage graphs & Split F/$dI$/$dW$; non-LIFO $dW$ & Zero state mismatches; exact restart & Evaluated system \\
\bottomrule
\end{tabularx}
\end{table*}

At PP=1, all four official Megatron capture modes complete with the same
six-significant-digit loss.  Under PP=2/VPP=2, eager and
\texttt{transformer\_engine} pass, whereas \texttt{local} fails at output-view
deallocation and \texttt{full\_iteration} fails on a second-step static-input
copy.  Dense delayed $dW$ is unavailable in this revision because it requires
the MoE expert-overlap path.  For the supported graph path, three paired trials
give 107.70 versus 51.05~ms at hidden size 2,048 and 228.45 versus 133.05~ms at
hidden size 4,096.  The median paired speedups are 2.191$\times$ and
1.751$\times$, respectively.  All 20
logged losses match their eager trajectories at the reported precision; these
ratios remain within Megatron.

Table~\ref{tab:correctness} summarizes the state checks.  Each 20-step run uses
native TE full backward as the reference for the corresponding \system{} eager
or captured schedule and crosses the four-entry amax-history rollover.  At each
microbatch action and optimizer step, we compare F/$dI$/$dW$ tensors, gradients,
parameters, AdamW state, FP8 scaling state, and cache contents.  We also check addresses,
epochs, and pending resources required by the contract.  All mismatch counts
are zero.  A fresh-process DCP run saves at step 3 and continues through step 8
for eager/captured $\times$ Interleaved/ZB-V.  Persistent state and subsequent
data batches match uninterrupted execution; process-local tokens and addresses
are rebuilt after restart.

\begin{table*}[t]
\centering
\caption{Numerical and runtime-state checks.  Each pipeline configuration
compares \system{} with native TE full backward under the same stage mapping.}
\label{tab:correctness}
\footnotesize
\setlength{\tabcolsep}{4pt}
\begin{tabularx}{\textwidth}{@{}>{\raggedright\arraybackslash}p{0.14\textwidth}>{\raggedright\arraybackslash}p{0.27\textwidth}>{\raggedright\arraybackslash}p{0.39\textwidth}>{\raggedright\arraybackslash}X@{}}
\toprule
Check & Compared executions & State compared & Finding \\
\midrule
Stream ordering & Fixed producer/consumer graphs; one-way versus two-way stream handoff & Output, parameters, and FP8-state/cache hashes & One-way path stale in all 600 checks (3 processes, 200 each); two-way path has none \\
Stepwise state & Hidden size 512, 20 steps; native TE versus \system{} eager and captured schedules & Per-microbatch F/$dI$/$dW$, gradients, AdamW, FP8 state/cache, versions, addresses, and pending resources & Zero mismatches in every category \\
Pipeline mapping & Hidden size 512, 20 steps; PP=2 and PP=4 captured Interleaved versus native TE under the same mapping & Loss, tensors, gradients, parameters, AdamW, FP8 state/cache, versions, retained-work state, addresses, and memory growth & Zero mismatches in all 28 categories at both PP degrees \\
Fresh-process resume & Save at step 3 and continue through step 8; eager/captured $\times$ Interleaved/ZB-V & Model, optimizer, learning-rate state, dataloader, FP8 state, versions, cache refresh, and pending work & Zero mismatches in all four configurations \\
Data-parallel correctness & PP=2$\times$DP=2 with eager/captured schedules and reducers & Peer gradients, parameters, runtime state, and completed resources & All equality checks hold \\
C4 training quality & 5,000 train plus 100 held-out steps; native versus \system{} & Held-out cross-entropy/perplexity, run variation, and unchanged validation state & Meets the 5\% perplexity non-inferiority criterion \\
H800 training stability & H800, 100 train plus 20 held-out steps; native TE, \system{} TE, and \system{} TorchAO & Finite state, training cross-entropy ratio, and held-out cross-entropy & All three trajectories remain finite and decreasing \\
\bottomrule
\end{tabularx}
\end{table*}

The contract also exposes invalid transitions at a precise host boundary.
Seven CPU-only injections cover the six violation families in
Table~\ref{tab:faults}; the process initializes no CUDA context.  Every
injection is rejected at its first invalid operation.  A rejected stale
reference preserves the current tape, incomplete work remains uncommitted, and
failed gradient-buffer or reduction versions require rebuilding the affected
runtime state.

\begin{table}[t]
\centering
\caption{Injected contract violations and their detection points.  The final
row contains separate pending-work and in-flight-reduction injections.}
\label{tab:faults}
\footnotesize
\setlength{\tabcolsep}{2.2pt}
\begin{tabularx}{\columnwidth}{@{}>{\raggedright\arraybackslash}p{0.35\columnwidth}>{\raggedright\arraybackslash}p{0.27\columnwidth}>{\raggedright\arraybackslash}X@{}}
\toprule
Injected violation & Detected at & Result \\
\midrule
Old retained-work reference after slot reuse & retained-work lookup & current record preserved \\
Repeated $dI$ consumption & ownership check & duplicate rejected \\
Finalize with one missing $dW$ & gradient-buffer finalization & optimizer version incomplete \\
Start a write after an aborted buffer write & gradient-buffer write start & version invalid; rebuild \\
Use a stale weight cache & forward version check & cache refresh required \\
Checkpoint with a live tape or reduction & checkpoint serialization & checkpoint withheld \\
\bottomrule
\end{tabularx}
\end{table}

Within each \system{}/native TE pair, cloned parameters and inputs allow exact
tensor fingerprints at every microbatch action and optimizer step.  The resume
test compares each restarted suffix with uninterrupted execution under the same
schedule.  Megatron's standard log exposes six significant digits of loss, so
its graph/eager trajectory is compared at that precision.

\system{} runtime state occupies only $23{,}969$ bytes within a
$278{,}484{,}866$-byte DCP checkpoint.  Four $5{,}677$-byte stage payloads and a
$1{,}261$-byte schema payload retain FP8 state, weight versions, and persistent
configuration.  The $5{,}000$-step C4 pretraining trajectory satisfies the 5\%
perplexity non-inferiority bound alongside matched run-to-run variation.  In the
H800 stability verification ($100$ train plus $20$ held-out steps), the
last-ten to first-ten training cross-entropy ratios decrease steadily to
$0.4765$, $0.4545$, and $0.4514$ for native TE, \system{} TE, and \system{}
TorchAO, respectively, with corresponding held-out cross-entropy means of
$3.3690$, $3.0013$, and $3.0062$.  All three configurations demonstrate robust
numerical convergence under low-precision execution.

\subsection{Complete-step Hopper performance}

Table~\ref{tab:hopper} reports an initial three-trial H800 set and an
independently launched two-trial replication set without pooling them.
Across TE
shapes, captured Interleaved is 1.884--2.792$\times$ faster than native eager;
captured ZB-V is 1.819--2.700$\times$ faster.  At hidden size 4,096, the TorchAO dynamic
Float8 output boundary reaches 1.710--1.744$\times$ for Interleaved
and 1.653--1.683$\times$ for ZB-V.  Schedule ordering depends on the workload:
ZB-V is faster in some smaller-microbatch cells, whereas Interleaved is faster
with eight microbatches at hidden size 4,096.

\begin{table*}[t]
\centering
\caption{Complete optimizer-step performance on two H800 PCIe GPUs.  Native
uses eager Interleaved 1F1B with full backward; \system{} candidates use
captured split backward under the named schedule.  Values are slowest-rank medians,
speedups are medians of paired per-trial ratios, and the initial and replication
trial sets remain separate.}
\label{tab:hopper}
\scriptsize
\setlength{\tabcolsep}{4.0pt}
\begin{tabular}{@{}llrrrrr@{}}
\toprule
Backend / hidden size & Trial set ($n$) & Native ms & Interleaved ms & Speedup & ZB-V ms & Speedup \\
\midrule
TE / 2,048 & Initial (3) & 44.887 & 16.184 & 2.792$\times$ & 16.648 & 2.696$\times$ \\
TE / 2,048 & Replication (2) & 44.727 & 16.391 & 2.729$\times$ & 16.564 & 2.700$\times$ \\
TE / 4,096 & Initial (3) & 89.469 & 47.329 & 1.895$\times$ & 49.025 & 1.824$\times$ \\
TE / 4,096 & Replication (2) & 89.138 & 47.319 & 1.884$\times$ & 49.002 & 1.819$\times$ \\
TorchAO / 4,096 & Initial (3) & 94.052 & 53.944 & 1.744$\times$ & 55.825 & 1.683$\times$ \\
TorchAO / 4,096 & Replication (2) & 92.315 & 53.995 & 1.710$\times$ & 55.862 & 1.653$\times$ \\
\bottomrule
\end{tabular}
\end{table*}

\begin{figure*}[t]
  \centering
  \includegraphics[width=\textwidth]{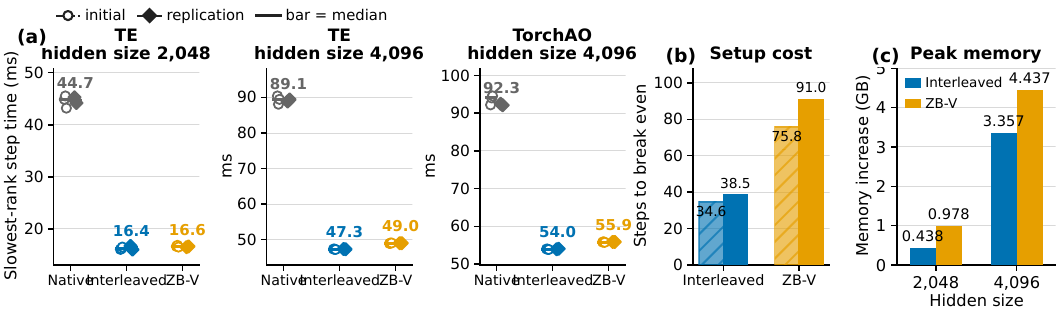}
  \caption{H800 complete-step performance and amortization.  (a) Workload-specific
  axes show every slowest-rank process trial and keep the initial trials (open
  circles, $n=3$) separate from the replication trials (filled diamonds,
  $n=2$); short bars mark
  trial-set medians.  (b) Hidden-size-4,096 break-even includes initialization and
  the first capture-heavy step.  (c) Peak allocation above native eager.}
  \label{fig:hopper}
\end{figure*}

Figure~\ref{fig:hopper} shows that setup costs change the usable regime.  At
hidden size 4,096, Interleaved reaches estimated total-step break-even after
34.55 steps in the initial trial set and 38.53 steps in the replication set.
ZB-V requires 75.77 and 91.02 steps.  At hidden sizes 2,048 and 4,096,
peak-memory increases are 0.438/3.357~GB for Interleaved and 0.978/4.437~GB for
ZB-V.  Each
steady rank owns 48 prebuilt graphs and performs 96 caller/graph handoffs per
step.  The measured benefit therefore applies to multi-step training after
these startup and memory costs.

\subsection{Causal attribution: direct destinations}

The five-configuration Ada matrix separates the action interface, graph
capture, and gradient destination.  Native eager takes 88.312~ms, \system{}
eager 104.392~ms, capture with copied gradients 99.134~ms, capture with direct
placement 87.107~ms, and captured ZB-V with direct placement 94.935~ms.  \system{} eager
is slower than native; capture recovers part of the gap, and
direct placement makes the otherwise matched path 1.138$\times$ faster.
The paired direct/copy break-even median is 125.70 total steps, its peak
allocation is 1,966,080 bytes lower, and neither path shows persistent memory
growth.

The same five-configuration decomposition on H800 uses five fresh processes per
configuration and is analyzed separately from the initial and replication
cohorts.  Native eager, \system{} eager, capture with copied gradients,
capture with direct placement, and captured ZB-V with direct placement take 94.582, 122.498,
54.126, 47.833, and 49.492~ms.  \system{} eager is slower than native.  Capture
with copied gradients reaches 1.747$\times$, and direct placement raises the
speedup to 1.977$\times$ while running 1.132$\times$ faster than the otherwise
matched copied-gradient path.  The direct path wins all five
copied-gradient/direct-placement pairs,
reduces step time by 11.67\%, and uses 1,900,544 fewer peak-allocation bytes.

Capture with copied gradients, capture with direct placement, and captured ZB-V
break even against native after 33.10, 30.17, and 77.99 total steps.  Their peak allocations are
3.359, 3.357, and 4.437~GB above native.  \system{} eager adds 3.949~GB and has
no break-even because its steady step is slower.  Sample coefficients of
variation range from 0.17\% to 1.45\%, and every configuration shows zero
persistent allocated-memory growth.  At this hidden-size-4,096,
eight-microbatch point, ZB-V is slightly slower than captured Interleaved.

Matched Nsight Systems traces identify the removed work.  Captured executions
retain graph launches, compute, point-to-point communication, optimizer work,
and stream events.  The complete Ada trace records 96 matrix-gradient copies
totaling 5,033,164,800 bytes and 12.320922~ms; the other Ada trace provides a
95-copy lower bound.  Separate H800 traces record 96 copies per rank with
copied gradients and zero under direct placement.  Their projected 4.71--4.79~GB volumes and
5.64~ms durations remain lower bounds because event collection is incomplete.
Profiler runs are excluded from timing results.

\begin{figure*}[t]
  \centering
  \includegraphics[width=\textwidth]{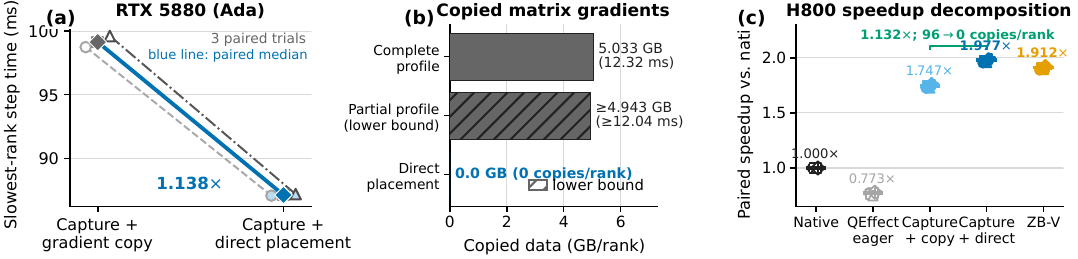}
  \caption{Why direct gradient placement helps.  (a) Three paired Ada trials
  compare capture with gradient copies against capture with direct placement.  (b) Direct placement
  removes the measured matrix-gradient copy projection; the hatched trace is a
  lower bound.  (c) Five matched H800 trials per configuration separate native,
  \system{} eager, capture with copied gradients, capture with direct placement,
  and captured ZB-V.  Short bars mark medians of the five paired speedup ratios.
  Separate Hopper traces record
  96-to-zero copies per rank.}
  \label{fig:attribution}
\end{figure*}

\subsection{How microbatch count changes resources}

\begin{figure*}[t]
  \centering
  \includegraphics[width=\textwidth]{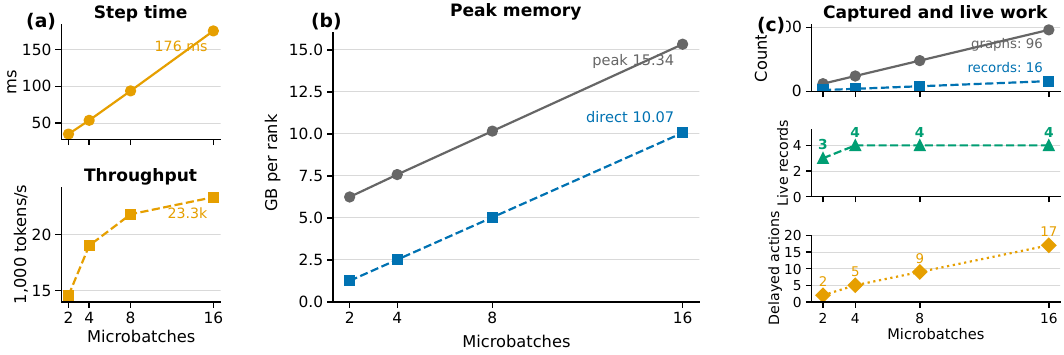}
  \caption{How configured microbatches affect time, memory, and live backward
  work in a ZB-V step on two RTX 5880 GPUs.
  (a) Step time and global throughput.  (b) Peak allocation and direct-gradient capacity.
  (c) Captured graphs and retained-work capacity scale linearly, whereas the
  number of live backward records saturates at four; delayed $dW$ count continues to
  grow.  Points are medians of three fresh processes per setting.}
  \label{fig:scaling}
\end{figure*}

Figure~\ref{fig:scaling} varies the configured microbatches in a ZB-V step.
Slowest-rank time rises from 35.124 to 175.589~ms across 2--16 microbatches,
while global throughput rises from 14.6k to 23.3k tokens/s.  Graph counts grow
from 12 to 96 per rank, and direct-gradient capacity grows from 1.258 to
10.066~GB.  Peak allocation rises from 6.249 to 15.341~GB with zero persistent
growth.

Capacity and schedule pressure diverge.  Although the graph and arena reserve
every fixed microbatch slot, peak live tapes saturate at 3/4/4/4 while delayed
$dW$ actions grow as 2/5/9/17.  Capture host time grows more sharply to
46.23/265.21/1593.84/3778.96~ms, making amortization part of configuration.

\subsection{Single-node pipeline mapping}

The fixed-work experiment keeps a four-stage model with hidden size 4,096, feed-forward
size 11,008, sequence length 256, and 2,048 tokens per step constant.  PP=2
places two stages per GPU; PP=4 places one.  The PP=4 mapping reduces step time
in all three pairs on both platforms,
with 1.366--1.370$\times$ speedup and 68.28--68.49\% strong-scaling efficiency
(Table~\ref{tab:pp-scaling}).

\begin{table}[t]
\centering
\caption{Single-node fixed-work pipeline scaling across three independent process pairs.
Times are cross-trial medians; speedup is the median paired ratio.}
\label{tab:pp-scaling}
\small
\setlength{\tabcolsep}{4pt}
\begin{tabular}{@{}lrrrr@{}}
\toprule
GPU platform & PP=2 ms & PP=4 ms & Speedup & Efficiency \\
\midrule
RTX 5880 Ada & 86.932 & 63.662 & 1.366$\times$ & 68.28\% \\
H800 PCIe & 47.593 & 34.730 & \textbf{1.370$\times$} & \textbf{68.49\%} \\
\bottomrule
\end{tabular}
\end{table}

Separate 20-step H800 runs compare PP=2 and PP=4 with native TE full backward
under the same mapping; all 28 mismatch categories are zero.  Comparing each
mapping with its matching reference preserves the legal PP-dependent order of
delayed-scaling effects.

The PP=2$\times$DP=2 run separately checks peer equality and reduction
completion under data parallelism.

\section{Discussion}
\label{sec:discussion}

\subsection{Effects and resources should remain distinct}

A token orders numerical updates that affect later quantization.  Retained
backward work also needs a clear owner and lifetime.  Caches need version
checks, gradient buffers need stable destinations and complete writes, and
stream events mark device completion.  Keeping these relations distinct
preserves pipeline concurrency and reports the exact failed relation.

\subsection{Schedule and resource tradeoffs}

\system{} allows delayed-scaling TE to follow ZB-V's non-LIFO $dW$ order.
Schedule choice remains workload dependent.  At hidden size 4,096 with eight
microbatches, Interleaved is about 1.7~ms faster and uses roughly 1.08~GB less
peak memory.  Smaller cells can favor ZB-V.  The schedule changes overlap,
live activations, delayed-$dW$ work, and graph handoffs.  Provisioned graphs and direct-gradient bytes, in contrast,
grow with $M$ even after live retained-work pressure saturates.  Reusing a
gradient slot would also have to preserve the \texttt{main\_grad} address bound into its captured $dW$
action.  Schedule selection and capacity reuse are therefore optimization
problems above the four correctness relations; GraCE studies the related
question of graph profitability~\cite{ghosh2026grace}.

\subsection{What a backend contract would need}

The four relations do not depend on TE class names.  A stateful backend needs
to expose its numerical transition order and the object retained between $dI$
and $dW$.  It must also reveal when a cached representation is current and
which state survives restart.  A backend with static scales or immediate
$dW$ can omit the relations it does not use.  Our TE 2.18 integration obtains
these hooks from its recipes, \texttt{WeightGradStore}, cache refresh, and fused
\texttt{main\_grad} interfaces without changing TE itself.

\subsection{Limitations}

The full delayed-scaling runtime is evaluated with TE; TorchAO exercises stage
boundaries without internal retained stores.  All performance benchmarks are
conducted on single-node PCIe systems (PP=2 and PP=4 Interleaved), leaving
multi-node InfiniBand clusters, tensor/expert parallelism, and PP=4 ZB-V for
future investigation.  Importantly, the four \system{} relations govern purely
intra-stage device semantics; because inter-node pipeline communication relies on
standard NCCL point-to-point primitives, stage-level CUDA Graph capture and
generational tape tracking remain strictly orthogonal to the underlying network
fabric.  Additionally, fixed-shape binding reserves one direct-gradient view per
configured microbatch, causing graph count and arena footprint to scale linearly
with $M$.  Checkpoints remain topology-preserving and are captured strictly at
quiescent optimizer boundaries.  Finally, partial profiler buffer wrap-arounds
render reported memory copy volumes lower bounds, although Hopper copy-event
counts are exact.

\section{Related Work}

\textbf{Effects in compiler intermediate representations.}
StableHLO uses opaque tokens to impose execution order, JAX distinguishes
compiler from runtime tokens, and MLIR associates effects with abstract
resources~\cite{openxla2026stablehlo,jax2024effects,mlir2026sideeffects}.
These abstractions order effects.  Split pipeline backward also needs to name
retained work whose consumer may arrive out of order.  The \system{} contract
combines order with ownership, version, and device completion for that case.

\textbf{Graph capture and compilation.}
PyTorch 2 captures Python programs and lowers tensor computation through
Dynamo, AOTAutograd, and Inductor~\cite{ansel2024pytorch2}; torch.fx,
Nimble, and TASO provide graph transformation, task scheduling, and substitution
search~\cite{reed2022torchfx,kwon2020nimble,jia2019taso}.  GraCE broadens CUDA
Graph coverage through pointer indirection and selective deployment
~\cite{ghosh2026grace}.  Our problem appears at the boundary between the
captured tensor program and a scheduler that reorders F/$dI$/$dW$ around
persistent backend state.

\textbf{Low-precision training.}
Mixed-precision and FP8 training combine higher-precision master state with
reduced-precision arithmetic~\cite{micikevicius2018mixed,micikevicius2022fp8,
peng2023fp8lm,perez2023fp8,balanca2024scalify}.  TE supports persistent FP8
buffers and CUDA Graph execution for documented Megatron schedules
~\cite{nvidia2026transformerenginegraphs}.  Delayed scaling gives those buffers
a temporal order and a cache version, while Static $\mu$nit Scaling removes
dynamic scale state through a different numerical design~\cite{narayan2025unit}.

\textbf{Pipeline and distributed runtimes.}
GPipe and PipeDream develop pipeline schedules and weight versioning
~\cite{huang2019gpipe,narayanan2019pipedream,narayanan2021pipedream2bw};
Megatron and DeepSpeed scale distributed transformer training
~\cite{shoeybi2019megatron,narayanan2021megatron,rasley2020deepspeed}.  Zero
Bubble separates $dI$ and $dW$, while JaxPP, Piper, and GraphPipe expand
schedule and topology support~\cite{qi2024zerobubble,qi2024controllable,
xhebraj2025jaxpp,frisella2025piper,jeon2025graphpipe}.  \system{} binds hidden
low-precision state and each retained $dW$ object to the resulting action order.

\textbf{Distributed graph optimization and persistence.}
Unity and Alpa jointly optimize computation, placement, and parallelization
~\cite{unger2022unity,zheng2022alpa}; Checkmate models rematerialization
lifetimes~\cite{jain2020checkmate}.  Their optimization problems differ from a
retained gradient whose later consumer and destination are fixed by capture.
PyTorch DCP provides storage~\cite{pytorch2026dcp}; the \system{} restart
boundary separates persistent numerical state from rebuilt process-local
resources.

\section{Conclusion}

CUDA Graph replay preserves addresses and tensor operations, but it does not
by itself preserve the meaning of hidden state or retained backward work.
\system{} makes four missing relations explicit: state order, resource
ownership, weight version, and device completion.  Eager and captured runs
match native TE full backward through delayed-scaling rollover.  All seven
injected violations are rejected at their first boundary, and fresh-process
restart reproduces uninterrupted execution.  On H800, captured TE runs are
1.82--2.79$\times$ faster than native eager.  Direct gradient placement removes
96 copies per rank and improves its matched path by 1.132$\times$.  Once the
four relations are explicit, capture can preserve pipeline concurrency without
confusing a reused address with the state or work it represents.

\section*{Acknowledgment}
OpenAI Codex~\cite{openai2026codex} assisted with manuscript organization and language editing.  The authors developed, evaluated, and verified all technical designs, implementations, and empirical results.

\bibliographystyle{IEEEtran}
\bibliography{references}

\begin{thebibliography}{10}
\providecommand{\url}[1]{#1}
\csname url@samestyle\endcsname
\providecommand{\newblock}{\relax}
\providecommand{\bibinfo}[2]{#2}
\providecommand{\BIBentrySTDinterwordspacing}{\spaceskip=0pt\relax}
\providecommand{\BIBentryALTinterwordstretchfactor}{4}
\providecommand{\BIBentryALTinterwordspacing}{\spaceskip=\fontdimen2\font plus
\BIBentryALTinterwordstretchfactor\fontdimen3\font minus
  \fontdimen4\font\relax}
\providecommand{\BIBforeignlanguage}[2]{{%
\expandafter\ifx\csname l@#1\endcsname\relax
\typeout{** WARNING: IEEEtran.bst: No hyphenation pattern has been}%
\typeout{** loaded for the language `#1'. Using the pattern for}%
\typeout{** the default language instead.}%
\else
\language=\csname l@#1\endcsname
\fi
#2}}
\providecommand{\BIBdecl}{\relax}
\BIBdecl

\bibitem{micikevicius2022fp8}
\BIBentryALTinterwordspacing
P.~Micikevicius, D.~Stosic, N.~Burgess, M.~Cornea, P.~Dubey, R.~Grisenthwaite,
  S.~Ha, A.~Heinecke, P.~Judd, J.~Kamalu, N.~Mellempudi, S.~F. Oberman,
  M.~Shoeybi, M.~Y. Siu, and H.~Wu, ``{FP8} formats for deep learning,''
  \emph{arXiv preprint arXiv:2209.05433}, 2022. [Online]. Available:
  \url{https://arxiv.org/abs/2209.05433}
\BIBentrySTDinterwordspacing

\bibitem{peng2023fp8lm}
\BIBentryALTinterwordspacing
H.~Peng, K.~Wu, Y.~Wei, G.~Zhao, Y.~Yang, Z.~Liu, Y.~Xiong, Z.~Yang, B.~Ni,
  J.~Hu, R.~Li, M.~Zhang, C.~Li, J.~Ning, R.~Wang, Z.~Zhang, S.~Liu, J.~Chau,
  H.~Hu, and P.~Cheng, ``{FP8-LM}: Training {FP8} large language models,''
  \emph{arXiv preprint arXiv:2310.18313}, 2023. [Online]. Available:
  \url{https://arxiv.org/abs/2310.18313}
\BIBentrySTDinterwordspacing

\bibitem{narayanan2021megatron}
\BIBentryALTinterwordspacing
D.~Narayanan, M.~Shoeybi, J.~Casper, P.~LeGresley, M.~Patwary, V.~A.
  Korthikanti, D.~Vainbrand, P.~Kashinkunti, J.~Bernauer, B.~Catanzaro,
  A.~Phanishayee, and M.~Zaharia, ``Efficient large-scale language model
  training on {GPU} clusters using {Megatron-LM},'' in \emph{Proceedings of the
  International Conference for High Performance Computing, Networking, Storage
  and Analysis}, 2021, pp. 1--15. [Online]. Available:
  \url{https://doi.org/10.1145/3458817.3476209}
\BIBentrySTDinterwordspacing

\bibitem{qi2024zerobubble}
\BIBentryALTinterwordspacing
P.~Qi, X.~Wan, G.~Huang, and M.~Lin, ``Zero bubble (almost) pipeline
  parallelism,'' in \emph{International Conference on Learning
  Representations}, 2024. [Online]. Available:
  \url{https://openreview.net/forum?id=tuzTN0eIO5}
\BIBentrySTDinterwordspacing

\bibitem{ansel2024pytorch2}
\BIBentryALTinterwordspacing
J.~Ansel, E.~Yang, H.~He, N.~Gimelshein, A.~Jain, M.~Voznesensky, B.~Bao,
  P.~Bell, D.~Berard, E.~Burovski, G.~Chauhan, A.~Chourdia, W.~Constable,
  A.~Desmaison, Z.~DeVito, E.~Ellison, W.~Feng, J.~Gong, M.~Gschwind, B.~Hirsh,
  S.~Huang, K.~Kalambarkar, L.~Kirsch, M.~Lazos, M.~Lezcano, Y.~Liang,
  J.~Liang, Y.~Lu, C.~K. Luk, B.~Maher, Y.~Pan, C.~Puhrsch, M.~Reso,
  M.~Saroufim, M.~Y. Siraichi, H.~Suk, S.~Zhang, M.~Suo, P.~Tillet, X.~Zhao,
  E.~Wang, K.~Zhou, R.~Zou, X.~Wang, A.~Mathews, W.~Wen, G.~Chanan, P.~Wu, and
  S.~Chintala, ``{PyTorch 2}: Faster machine learning through dynamic python
  bytecode transformation and graph compilation,'' in \emph{Proceedings of the
  29th ACM International Conference on Architectural Support for Programming
  Languages and Operating Systems, Volume 2}, 2024, pp. 929--947. [Online].
  Available: \url{https://doi.org/10.1145/3620665.3640366}
\BIBentrySTDinterwordspacing

\bibitem{nvidia2026cudagraphs}
\BIBentryALTinterwordspacing
{NVIDIA Corporation}, ``{CUDA} programming guide: {CUDA} graphs,'' 2026,
  accessed 2026-08-23. [Online]. Available:
  \url{https://docs.nvidia.com/cuda/cuda-programming-guide/04-special-topics/cuda-graphs.html}
\BIBentrySTDinterwordspacing

\bibitem{ghosh2026grace}
\BIBentryALTinterwordspacing
A.~Ghosh, A.~Nayak, A.~Panwar, and A.~Basu, ``{GraCE}: Unlocking {CUDA} graphs
  with compiler support for {ML} workloads,'' in \emph{20th USENIX Symposium on
  Operating Systems Design and Implementation}.\hskip 1em plus 0.5em minus
  0.4em\relax USENIX Association, 2026, pp. 1927--1947. [Online]. Available:
  \url{https://www.usenix.org/conference/osdi26/presentation/ghosh}
\BIBentrySTDinterwordspacing

\bibitem{openxla2026stablehlo}
\BIBentryALTinterwordspacing
{OpenXLA Project}, ``{StableHLO} specification,'' 2026, accessed 2026-08-29.
  [Online]. Available: \url{https://openxla.org/stablehlo/spec}
\BIBentrySTDinterwordspacing

\bibitem{jax2024effects}
\BIBentryALTinterwordspacing
{JAX Authors}, ``Sequencing side-effects in {JAX},'' 2024, jAX Enhancement
  Proposal 10657; accessed 2026-08-29. [Online]. Available:
  \url{https://docs.jax.dev/en/latest/jep/10657-sequencing-effects.html}
\BIBentrySTDinterwordspacing

\bibitem{mlir2026sideeffects}
\BIBentryALTinterwordspacing
{MLIR Project}, ``Side effects and speculation,'' 2026, accessed 2026-08-29.
  [Online]. Available:
  \url{https://mlir.llvm.org/docs/Rationale/SideEffectsAndSpeculation/}
\BIBentrySTDinterwordspacing

\bibitem{nvidia2026transformerenginegraphs}
\BIBentryALTinterwordspacing
{NVIDIA Corporation}, ``Transformer engine and {Megatron-LM} {CUDA} graph
  support,'' CUDA Graph Best Practice for PyTorch, 2026, accessed 2026-08-23.
  [Online]. Available:
  \url{https://docs.nvidia.com/dl-cuda-graph/latest/torch-cuda-graph/te-megatron-cuda-graphs.html}
\BIBentrySTDinterwordspacing

\bibitem{narayan2025unit}
\BIBentryALTinterwordspacing
S.~Narayan, A.~Gupta, M.~Paul, and D.~Blalock, ``{$\mu$}nit scaling: Simple and
  scalable {FP8} {LLM} training,'' in \emph{Proceedings of the 42nd
  International Conference on Machine Learning}, ser. Proceedings of Machine
  Learning Research, vol. 267.\hskip 1em plus 0.5em minus 0.4em\relax PMLR,
  2025, pp. 45\,720--45\,736. [Online]. Available:
  \url{https://proceedings.mlr.press/v267/narayan25b.html}
\BIBentrySTDinterwordspacing

\bibitem{liang2025torchtitan}
\BIBentryALTinterwordspacing
W.~Liang, T.~Liu, L.~Wright, W.~Constable, A.~Gu, C.-C. Huang, I.~Zhang,
  W.~Feng, H.~Huang, J.~Wang, S.~Purandare, G.~Nadathur, and S.~Idreos,
  ``{TorchTitan}: One-stop {PyTorch} native solution for production ready {LLM}
  pretraining,'' in \emph{International Conference on Learning
  Representations}, 2025. [Online]. Available:
  \url{https://openreview.net/forum?id=SFN6Wm7YBI}
\BIBentrySTDinterwordspacing

\bibitem{pytorch2026dcp}
\BIBentryALTinterwordspacing
{PyTorch Contributors}, ``Distributed checkpoint,'' 2026, accessed 2026-08-23.
  [Online]. Available:
  \url{https://docs.pytorch.org/docs/main/distributed.checkpoint.html}
\BIBentrySTDinterwordspacing

\bibitem{raffel2020t5}
\BIBentryALTinterwordspacing
C.~Raffel, N.~Shazeer, A.~Roberts, K.~Lee, S.~Narang, M.~Matena, Y.~Zhou,
  W.~Li, and P.~J. Liu, ``Exploring the limits of transfer learning with a
  unified text-to-text transformer,'' \emph{Journal of Machine Learning
  Research}, vol.~21, no. 140, pp. 1--67, 2020. [Online]. Available:
  \url{https://jmlr.org/papers/v21/20-074.html}
\BIBentrySTDinterwordspacing

\bibitem{reed2022torchfx}
\BIBentryALTinterwordspacing
J.~K. Reed, Z.~DeVito, H.~He, A.~Ussery, and J.~Ansel, ``{torch.fx}: Practical
  program capture and transformation for deep learning in python,'' in
  \emph{Proceedings of Machine Learning and Systems}, vol.~4, 2022. [Online].
  Available:
  \url{https://proceedings.mlsys.org/paper/2022/hash/7c98f9c7ab2df90911da23f9ce72ed6e-Abstract.html}
\BIBentrySTDinterwordspacing

\bibitem{kwon2020nimble}
\BIBentryALTinterwordspacing
W.~Kwon, G.-I. Yu, E.~Jeong, and B.-G. Chun, ``Nimble: Lightweight and parallel
  {GPU} task scheduling for deep learning,'' in \emph{Advances in Neural
  Information Processing Systems}, vol.~33, 2020. [Online]. Available:
  \url{https://papers.nips.cc/paper/2020/hash/5f0ad4db43d8723d18169b2e4817a160-Abstract.html}
\BIBentrySTDinterwordspacing

\bibitem{jia2019taso}
\BIBentryALTinterwordspacing
Z.~Jia, O.~Padon, J.~Thomas, T.~Warszawski, M.~Zaharia, and A.~Aiken, ``{TASO}:
  Optimizing deep learning computation with automatic generation of graph
  substitutions,'' in \emph{Proceedings of the 27th ACM Symposium on Operating
  Systems Principles}, 2019, pp. 47--62. [Online]. Available:
  \url{https://doi.org/10.1145/3341301.3359630}
\BIBentrySTDinterwordspacing

\bibitem{micikevicius2018mixed}
\BIBentryALTinterwordspacing
P.~Micikevicius, S.~Narang, J.~Alben, G.~Diamos, E.~Elsen, D.~Garcia,
  B.~Ginsburg, M.~Houston, O.~Kuchaiev, G.~Venkatesh, and H.~Wu, ``Mixed
  precision training,'' in \emph{International Conference on Learning
  Representations}, 2018. [Online]. Available:
  \url{https://openreview.net/forum?id=r1gs9JgRZ}
\BIBentrySTDinterwordspacing

\bibitem{perez2023fp8}
\BIBentryALTinterwordspacing
S.~P. Perez, Y.~Zhang, J.~Briggs, C.~Blake, J.~Levy-Kramer, P.~Balanca,
  C.~Luschi, S.~Barlow, and A.~W. Fitzgibbon, ``Training and inference of large
  language models using 8-bit floating point,'' in \emph{Workshop on Advancing
  Neural Network Training at NeurIPS}, 2023. [Online]. Available:
  \url{https://neurips.cc/virtual/2023/80694}
\BIBentrySTDinterwordspacing

\bibitem{balanca2024scalify}
\BIBentryALTinterwordspacing
P.~Balanca, S.~Hosegood, C.~Luschi, and A.~Fitzgibbon, ``Scalify: Scale
  propagation for efficient low-precision {LLM} training,'' in \emph{Workshop
  on Advancing Neural Network Training at ICML}, 2024. [Online]. Available:
  \url{https://openreview.net/forum?id=4IWCHWlb6K}
\BIBentrySTDinterwordspacing

\bibitem{huang2019gpipe}
\BIBentryALTinterwordspacing
Y.~Huang, Y.~Cheng, A.~Bapna, O.~Firat, D.~Chen, M.~X. Chen, H.~Lee, J.~Ngiam,
  Q.~V. Le, Y.~Wu, and Z.~Chen, ``{GPipe}: Efficient training of giant neural
  networks using pipeline parallelism,'' in \emph{Advances in Neural
  Information Processing Systems}, vol.~32, 2019. [Online]. Available:
  \url{https://papers.nips.cc/paper/2019/hash/093f65e080a295f8076b1c5722a46aa2-Abstract.html}
\BIBentrySTDinterwordspacing

\bibitem{narayanan2019pipedream}
\BIBentryALTinterwordspacing
D.~Narayanan, A.~Harlap, A.~Phanishayee, V.~Seshadri, N.~R. Devanur, G.~R.
  Ganger, P.~B. Gibbons, and M.~Zaharia, ``{PipeDream}: Generalized pipeline
  parallelism for {DNN} training,'' in \emph{Proceedings of the 27th ACM
  Symposium on Operating Systems Principles}, 2019, pp. 1--15. [Online].
  Available: \url{https://doi.org/10.1145/3341301.3359646}
\BIBentrySTDinterwordspacing

\bibitem{narayanan2021pipedream2bw}
\BIBentryALTinterwordspacing
D.~Narayanan, A.~Phanishayee, K.~Shi, X.~Chen, and M.~Zaharia,
  ``Memory-efficient pipeline-parallel {DNN} training,'' in \emph{Proceedings
  of the 38th International Conference on Machine Learning}, ser. Proceedings
  of Machine Learning Research, vol. 139.\hskip 1em plus 0.5em minus
  0.4em\relax PMLR, 2021, pp. 7937--7947. [Online]. Available:
  \url{https://proceedings.mlr.press/v139/narayanan21a.html}
\BIBentrySTDinterwordspacing

\bibitem{shoeybi2019megatron}
\BIBentryALTinterwordspacing
M.~Shoeybi, M.~Patwary, R.~Puri, P.~LeGresley, J.~Casper, and B.~Catanzaro,
  ``{Megatron-LM}: Training multi-billion parameter language models using model
  parallelism,'' \emph{arXiv preprint arXiv:1909.08053}, 2019. [Online].
  Available: \url{https://arxiv.org/abs/1909.08053}
\BIBentrySTDinterwordspacing

\bibitem{rasley2020deepspeed}
\BIBentryALTinterwordspacing
J.~Rasley, S.~Rajbhandari, O.~Ruwase, and Y.~He, ``{DeepSpeed}: System
  optimizations enable training deep learning models with over 100 billion
  parameters,'' in \emph{Proceedings of the 26th ACM SIGKDD International
  Conference on Knowledge Discovery and Data Mining}, 2020, pp. 3505--3506.
  [Online]. Available: \url{https://doi.org/10.1145/3394486.3406703}
\BIBentrySTDinterwordspacing

\bibitem{qi2024controllable}
\BIBentryALTinterwordspacing
P.~Qi, X.~Wan, N.~Amar, and M.~Lin, ``Pipeline parallelism with controllable
  memory,'' in \emph{Advances in Neural Information Processing Systems},
  vol.~37, 2024. [Online]. Available:
  \url{https://proceedings.neurips.cc/paper_files/paper/2024/hash/527dad0b9159805289906d5740a0bdd3-Abstract-Conference.html}
\BIBentrySTDinterwordspacing

\bibitem{xhebraj2025jaxpp}
\BIBentryALTinterwordspacing
A.~Xhebraj, S.~Lee, H.~Chen, and V.~Grover, ``Scaling deep learning training
  with {MPMD} pipeline parallelism,'' in \emph{Proceedings of Machine Learning
  and Systems}, vol.~7, 2025. [Online]. Available:
  \url{https://proceedings.mlsys.org/paper_files/paper/2025/hash/9f73d65a4186198152357be871345771-Abstract-Conference.html}
\BIBentrySTDinterwordspacing

\bibitem{frisella2025piper}
\BIBentryALTinterwordspacing
M.~Frisella, A.~Oentoro, X.~Gao, G.~Bernstein, and S.~Wang, ``{Piper}: Towards
  flexible pipeline parallelism for {PyTorch},'' in \emph{Practical Adoption
  Challenges of Machine Learning for Systems}, 2025, pp. 1--6. [Online].
  Available: \url{https://doi.org/10.1145/3766882.3767187}
\BIBentrySTDinterwordspacing

\bibitem{jeon2025graphpipe}
\BIBentryALTinterwordspacing
B.~Jeon, M.~Wu, S.~Cao, S.~Kim, S.~Park, N.~Aggarwal, C.~Unger, D.~Arfeen,
  P.~Liao, X.~Miao, M.~Alizadeh, G.~R. Ganger, T.~Chen, and Z.~Jia,
  ``{GraphPipe}: Improving performance and scalability of {DNN} training with
  graph pipeline parallelism,'' in \emph{Proceedings of the 30th ACM
  International Conference on Architectural Support for Programming Languages
  and Operating Systems, Volume 1}, 2025, pp. 557--571. [Online]. Available:
  \url{https://doi.org/10.1145/3669940.3707220}
\BIBentrySTDinterwordspacing

\bibitem{unger2022unity}
\BIBentryALTinterwordspacing
C.~Unger, Z.~Jia, W.~Wu, S.~Lin, M.~Baines, C.~E.~Q. Narvaez,
  V.~Ramakrishnaiah, N.~Prajapati, P.~McCormick, J.~Mohd-Yusof, X.~Luo,
  D.~Mudigere, J.~Park, M.~Smelyanskiy, and A.~Aiken, ``Unity: Accelerating
  {DNN} training through joint optimization of algebraic transformations and
  parallelization,'' in \emph{16th USENIX Symposium on Operating Systems Design
  and Implementation}.\hskip 1em plus 0.5em minus 0.4em\relax USENIX
  Association, 2022, pp. 267--284. [Online]. Available:
  \url{https://www.usenix.org/conference/osdi22/presentation/unger}
\BIBentrySTDinterwordspacing

\bibitem{zheng2022alpa}
\BIBentryALTinterwordspacing
L.~Zheng, Z.~Li, H.~Zhang, Y.~Zhuang, Z.~Chen, Y.~Huang, Y.~Wang, Y.~Xu,
  D.~Zhuo, E.~P. Xing, J.~E. Gonzalez, and I.~Stoica, ``Alpa: Automating inter-
  and intra-operator parallelism for distributed deep learning,'' in \emph{16th
  USENIX Symposium on Operating Systems Design and Implementation}.\hskip 1em
  plus 0.5em minus 0.4em\relax USENIX Association, 2022, pp. 559--578.
  [Online]. Available:
  \url{https://www.usenix.org/conference/osdi22/presentation/zheng-lianmin}
\BIBentrySTDinterwordspacing

\bibitem{jain2020checkmate}
\BIBentryALTinterwordspacing
P.~Jain, A.~Jain, A.~Nrusimha, A.~Gholami, P.~Abbeel, K.~Keutzer, I.~Stoica,
  and J.~E. Gonzalez, ``Checkmate: Breaking the memory wall with optimal tensor
  rematerialization,'' in \emph{Proceedings of Machine Learning and Systems},
  vol.~2, 2020, pp. 497--511. [Online]. Available:
  \url{https://proceedings.mlsys.org/paper/2020/hash/0b816ae8f06f8dd3543dc3d9ef196cab-Abstract.html}
\BIBentrySTDinterwordspacing

\bibitem{openai2026codex}
\BIBentryALTinterwordspacing
{OpenAI}, ``Codex documentation,'' 2026, accessed 2026-08-24. [Online].
  Available: \url{https://learn.chatgpt.com/codex/overview}
\BIBentrySTDinterwordspacing

\end{thebibliography}

\end{document}